\documentclass[ reprint,
 amsmath,amssymb,
 aps,
 prl
]{revtex4-2}
\usepackage[utf8]{inputenc}
\usepackage{amsmath}
\usepackage{xcolor}
\usepackage{graphicx}
\usepackage[colorlinks=true,allcolors=blue]{hyperref}
\usepackage{gensymb}
\usepackage{physics}
\usepackage{siunitx}

\newcommand{\NaRb}{$^{23}$Na$^{87}$Rb}

\newcommand{\upstate}{\ket{\uparrow}}
\newcommand{\downstate}{\ket{\downarrow}}

\usepackage[normalem]{ulem}

\begin{document}
\title{Probing site-resolved correlations in a spin system of ultracold molecules}
\author{Lysander Christakis}
\thanks{These authors contributed equally to this work.}
\affiliation{Department of Physics, Princeton University, Princeton, New Jersey 08544, USA}
\author{Jason S. Rosenberg}
\thanks{These authors contributed equally to this work.}
\affiliation{Department of Physics, Princeton University, Princeton, New Jersey 08544, USA}
\author{Ravin Raj}
\affiliation{Department of Physics, Princeton University, Princeton, New Jersey 08544, USA}
\author{Sungjae Chi}
\affiliation{Department of Physics, Princeton University, Princeton, New Jersey 08544, USA}
\author{Alan Morningstar}
\affiliation{Department of Physics, Princeton University, Princeton, New Jersey 08544, USA}
\author{David A. Huse}
\affiliation{Department of Physics, Princeton University, Princeton, New Jersey 08544, USA}
\author{Zoe Z. Yan}
\affiliation{Department of Physics, Princeton University, Princeton, New Jersey 08544, USA}
\author{Waseem S. Bakr}
\altaffiliation{Email: wbakr@princeton.edu}
\affiliation{Department of Physics, Princeton University, Princeton, New Jersey 08544, USA}
\date{\today}

\begin{abstract}

Synthetic quantum systems with interacting constituents play an important role in quantum information processing and in elucidating fundamental phenomena in many-body physics. Following impressive advances in cooling and trapping techniques, ensembles of ultracold polar molecules have emerged as a promising synthetic system that combines several advantageous properties~\cite{ni2008high,barry2014magneto,demarco2019degenerate,anderegg2019optical,cairncross2021,son2020collisional,schindewolf2022evaporation,vilas2022magneto,bohn2017cold,gadway2016strongly,moses2017new}. These include a large set of internal states for encoding quantum information, long nuclear and rotational coherence times~\cite{park2017second,seesselberg2018extending,caldwell2020,blackmore2021robust,burchesky2021rotational,lin2022} and long-range, anisotropic interactions. The latter are expected to allow the exploration of intriguing phases of correlated quantum matter, such as topological superfluids~\cite{cooper2009stable}, quantum spin liquids~\cite{yao2018quantum}, fractional Chern insulators~\cite{yao2013realizing} and quantum magnets~\cite{hazzard2013far,gorshkov2011}. 
Probing correlations in these phases is crucial to understand their microscopic properties, necessitating the development of new experimental techniques.
Here we use quantum gas microscopy~\cite{gross2021quantum} to measure the site-resolved dynamics of quantum correlations in a gas of polar molecules in a two-dimensional optical lattice. Using two rotational states of the molecules, we realize a spin-1/2 system where the particles are coupled via dipolar interactions, producing a quantum spin-exchange model~\cite{barnett2006quantum,gorshkov2011,hazzard2013far,yan2013observation}. Starting with the synthetic spin system prepared far from equilibrium, we study the evolution of correlations during the thermalization process for both spatially isotropic and anisotropic interactions. Furthermore, we study the correlation dynamics in a spin-anisotropic Heisenberg model engineered from the native spin-exchange model using Floquet techniques~\cite{choi2020robust,geier2021floquet,scholl2022}. These experiments push the frontier of probing and controlling interacting systems of ultracold molecules, with prospects for exploring new regimes of quantum matter and characterizing entangled states useful for quantum computation~\cite{DeMille2002,ni2018dipolar} and metrology~\cite{perlin2020spin}.
\end{abstract}

\maketitle
 
\begin{figure}[t]
\includegraphics[width=\columnwidth]{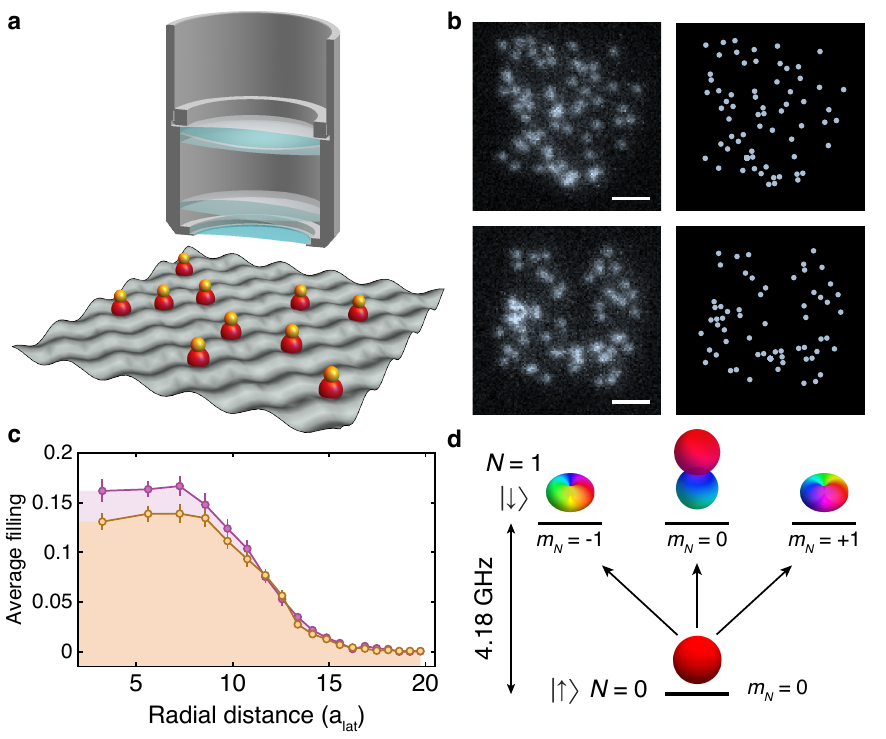}
\caption{\label{fig:qgm} \textbf{Molecular quantum gas microscope.} \textbf{a}, Schematic of the molecular quantum gas microscope. A high numerical-aperture objective is used to resolve the positions of individual molecules confined to a 2D optical lattice. \textbf{b}, Representative fluorescence images of Feshbach molecules (top left) and ground state molecules (bottom left) taken with the microscope and positions recorded with a reconstruction algorithm (top right and bottom right, respectively). Scale bars represent \SI{5}{\micro\meter}. \textbf{c}, Average density profile of a cloud of Feshbach molecules (purple) and the corresponding profile of detected molecules after round-trip STIRAP (orange). Error bars are s.e.m. \textbf{d}, Schematic of the rotational states used in the experiment, neglecting hyperfine structure.  The degeneracy of the rotational states is lifted by coupling between nuclear and rotational degrees of freedom (not shown, see Methods).  Our pseudospin states are chosen to be $\ket{N,m_N}=\ket{0,0}$ and an excited rotational state with predominant $\ket{1,-1}$ character.}
\end{figure}

Ultracold dipolar particles trapped in optical arrays have emerged as a powerful and flexible platform to probe idealized models of quantum many-body physics with long-range interactions~\cite{baranov2012condensed}. For example, studies using polar molecules~\cite{bohn2017cold,gadway2016strongly,moses2017new}, Rydberg atoms~\cite{browaeys2020many}, and magnetic atoms~\cite{chomaz2022dipolar} have explored a host of quantum phenomena including quantum magnetism~\cite{dePaz2013,yan2013observation, scholl2022}, symmetry-protected topological phases~\cite{deleseleuc2019observation} and extended Hubbard models~\cite{baier2016extended}. The platform of polar molecules in particular has several unique properties ~\cite{bohn2017cold,gadway2016strongly,moses2017new}. Ultracold polar molecules possess a tunable dipole moment in their electronic ground state, enabling strong, long-range interactions compatible with negligible population relaxation over experimental timescales. In addition, compared to atoms, molecules have new rotational and vibrational degrees of freedom that can be harnessed to store information in a large local Hilbert space and to engineer a rich set of many-body Hamiltonians. However, this complex internal structure also presents its own experimental challenges in the preparation and detection of quantum-state-controlled ensembles of molecules.

Two complementary approaches have been developed to address these challenges. On the one hand, the assembly of heteronuclear molecules from evaporatively cooled atomic quantum gases~\cite{ni2008high} has generated bulk ensembles with high phase space density, paving the path for the recent creation of degenerate molecular Fermi gases whose collective properties can be characterized with absorption imaging techniques~\cite{demarco2019degenerate,schindewolf2022evaporation}.  On the other hand, molecules have been prepared in optical tweezers~\cite{Kaufman2021}, realizing a bottom-up approach to controlling small numbers of molecules with single-molecule detection capabilities~\cite{anderegg2019optical,cairncross2021}. Here we demonstrate quantum gas microscopy of polar \NaRb~molecules in an optical lattice, combining features of both approaches: large numbers of molecules in their hyperfine, rovibronic, and motional ground state, coupled with single-molecule detection.  Quantum gas microscopy enables high-fidelity and simultaneous optical detection of ensembles of particles in a regime of low temperatures, high density, and strong interactions. Pioneered with atomic gases, this technique has enabled unprecedented local observations of quantum phase transitions, spin and charge correlations in Hubbard systems, and impurity physics~\cite{gross2021quantum}. Recently, we have extended microscopy techniques to quantum gases of non-interacting excited-state molecules, observing bosonic bunching correlations in their density fluctuations~\cite{rosenberg2021observation}.

In this work, we apply our ability to measure microscopic correlations to study out-of-equilibrium spin systems realized with interacting molecules. To that end, we prepare the molecules in their rovibrational ground state, which possesses a large body-frame electric dipole moment. We encode a spin-1/2 in the molecules' ground and first excited rotational states. An effective spin interaction arises from the resonant exchange of rotational excitations between pairs of molecules, as was demonstrated in previous work using polar $^{40}$K$^{87}$Rb molecules~\cite{yan2013observation, tobias2022reactions}.
Here, with molecules pinned in a deep two-dimensional optical lattice, the system realizes the quantum XY spin-exchange Hamiltonian~\cite{barnett2006quantum,gorshkov2011,hazzard2013far}
\begin{equation}
    H_{\rm XY} = \sum_{i> j} V(\mathbf{r}_i-\mathbf{r}_j) \left(S^x_iS^x_j + S^y_iS^y_j\right)
    \label{eq1}
\end{equation}
where $V(\mathbf{a})=J\langle(1-3\cos^2{\theta)}/|\mathbf{a}|^3\rangle$, $J$ characterizes the strength of the spin-exchange interaction, $S^{x(y)}_{i} = \sigma^{x(y)}_i/2$ are spin-1/2 operators for molecule $i$, $\mathbf{r}_i$ is the position of molecule $i$ measured in units of the lattice constant $a_\mathrm{lat}$, and $\theta$ is the angle between the quantization axis and the vector $\mathbf{a}$. The quantum average $\langle.\rangle$ accounts for the finite size of the molecule center-of-mass wavefunctions. XY Hamiltonians are some of the most extensively studied models for magnetism and excitation transport. However, like most quantum spin models, there remain many open questions in computationally intractable regimes, such as the prediction of low-temperature phases in the presence of frustration or the long-time dynamics of disordered or driven systems. Polar molecules are poised to be a leading platform to simulate such spin models due to the large ratio of spin coherence times to the interaction timescale $h/|J|$, as we demonstrate in this work. We observe the effects of the dipolar interactions by performing a series of quench dynamics experiments, where we initialize the molecules in a product state and then evolve under the interaction Hamiltonian in equation~(\ref{eq1}), after which we use a molecular quantum gas microscope to measure the correlations between the molecules with single-site resolution. For a lattice with low filling, the correlations exhibit weakly-damped oscillations with a frequency that depends on the site displacement, directly reflecting the long-range and anisotropic character of the interactions.

\section{Molecule preparation and detection}
We start our experiments by preparing an array of ground-state NaRb molecules~\cite{guo2016creation} in a 2D optical lattice (Fig.~\hyperref[fig:qgm]{1a}). The molecules are formed by loading degenerate Bose gases of Na and Rb into the lattice and using magnetoassociation followed by stimulated Raman adiabatic passage (STIRAP) to convert atom pairs into molecules in their rovibrational and hyperfine ground state~\cite{reichsollner2017quantum,moses2015creation}.  The lattice is sufficiently deep such that tunneling of molecules between lattice sites is negligible over the timescale of the experiments. The maximum achieved filling of ground-state molecules in the lattice is 15(1)\%, taking into account the detection efficiency (Fig.~\hyperref[fig:qgm]{1c}). To simulate an effective spin-1/2 system, we use microwaves near \SI{4.18}{\giga\hertz} to drive transitions between a hyperfine state in the ground manifold ($N=0$) and a hyperfine state in the first excited rotational manifold ($N=1$), labeled $\upstate$ and $\downstate$, respectively. 
For imaging, we reverse the STIRAP process, transferring molecules in $\upstate$ to the weakly bound Feshbach state, which we then dissociate. We detect the resulting Rb atoms with fluorescence imaging, allowing us to extract the position of each molecule in $\upstate$ with single-site resolution~\cite{rosenberg2021observation}. We do not detect the molecules in $\downstate$, so our current measurements do not distinguish between that state and an empty site, although this may be achieved in future work with bilayer techniques used for spin-resolved imaging in atomic microscopes~\cite{koepsell2020bilayer, yan2022two}.

\section{Rotational coherence}
\begin{figure}[th]
\includegraphics[width=\columnwidth]{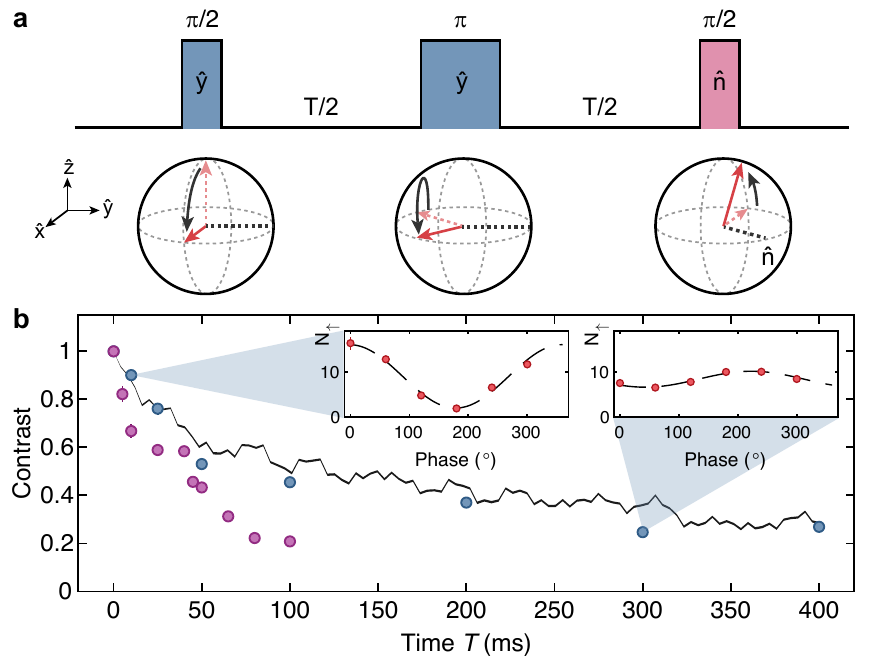}
\caption{\label{fig:coherence} \textbf{Rotational state coherence.} \textbf{a}, Microwave pulse sequence for the Ramsey coherence measurement including the spin echo $\pi$-pulse. \textbf{b}, Ramsey fringe contrast versus total precession time, with (blue) and without (purple) the spin echo pulse in the middle of the precession time, for a sample with peak filling of $1.0(2)\%$. The black line shows the predicted dynamics from exact diagonalization. Inset: representative Ramsey fringes from \SI{10}{\milli\second} and \SI{300}{\milli\second} precession times with a spin echo. Error bars are s.e.m.}
\end{figure}

An important benchmark for our quantum simulator is the coherence of the two-level system encoded in the rotational states of the molecules. A dominant effect limiting rotational coherence in ultracold molecules is the differential ac polarizability of the $N=0$ and $N=1$ states, leading to differential energy shifts due to the trapping light~\cite{Kotochigova2010,neyenhuis2012anisotropic,seesselberg2018extending,lin2021anisotropic,burchesky2021rotational,blackmore2018ultracold}.  
To address this issue, we apply an external magnetic field tuned to a near-magic condition where hyperfine effects nearly cancel the differential polarizability to within 1\% (see Methods). 

We probe the coherence of our system in a Ramsey experiment. Starting with all of the molecules in $\upstate$, we apply two $\pi/2$-pulses separated by a total free precession time $T$ (Fig.~\hyperref[fig:coherence]{2a}). A spin echo $\pi$-pulse can be inserted in the middle of the precession time, which mitigates dephasing due to quasi-static sources of single-particle decoherence but not dipolar interactions. The phase $\phi$ of the final $\pi/2$-pulse is scanned and the number of molecules in $\upstate$ is recorded. To reduce the possible influence of dipolar interactions between the molecules, we dilute the sample to have a peak molecule filling of 1.0(2)\%. For each precession time $T$, we fit the measured molecule number $N_{\uparrow}$ (Fig.~\hyperref[fig:coherence]{2b} inset) to $N_{\uparrow} = A\cos(\phi+\phi_0)+B$ to extract the contrast $A/B$. The results of this experiment are shown in Fig.~\ref{fig:coherence}. Without a spin echo pulse, the observed dephasing ($1/e$ time of \SI{56(2)}{\milli\second}) is faster than that predicted for an interacting XY spin system described by equation~(\ref{eq1}), most likely due to residual differential light shifts. Nevertheless, this decay time is much longer than the millisecond-scale nearest-neighbor interaction time, which is a favorable regime for future experiments studying spin-squeezing or preparing rotational superfluids~\cite{perlin2020spin,kwasigroch2014bose}. With the addition of a single spin echo $\pi$-pulse, we observe that the decay of the experimental contrast more closely tracks the dynamics expected from equation~(\ref{eq1}) out to $\SI{400}{\milli\second}$. This indicates that despite the diluteness of the system, the loss of contrast in this case is predominantly due to coherent interactions between spins rather than external sources of decoherence.

\section{Correlation dynamics in an XY spin model}
\begin{figure*}[tb]
\includegraphics[width=\textwidth]{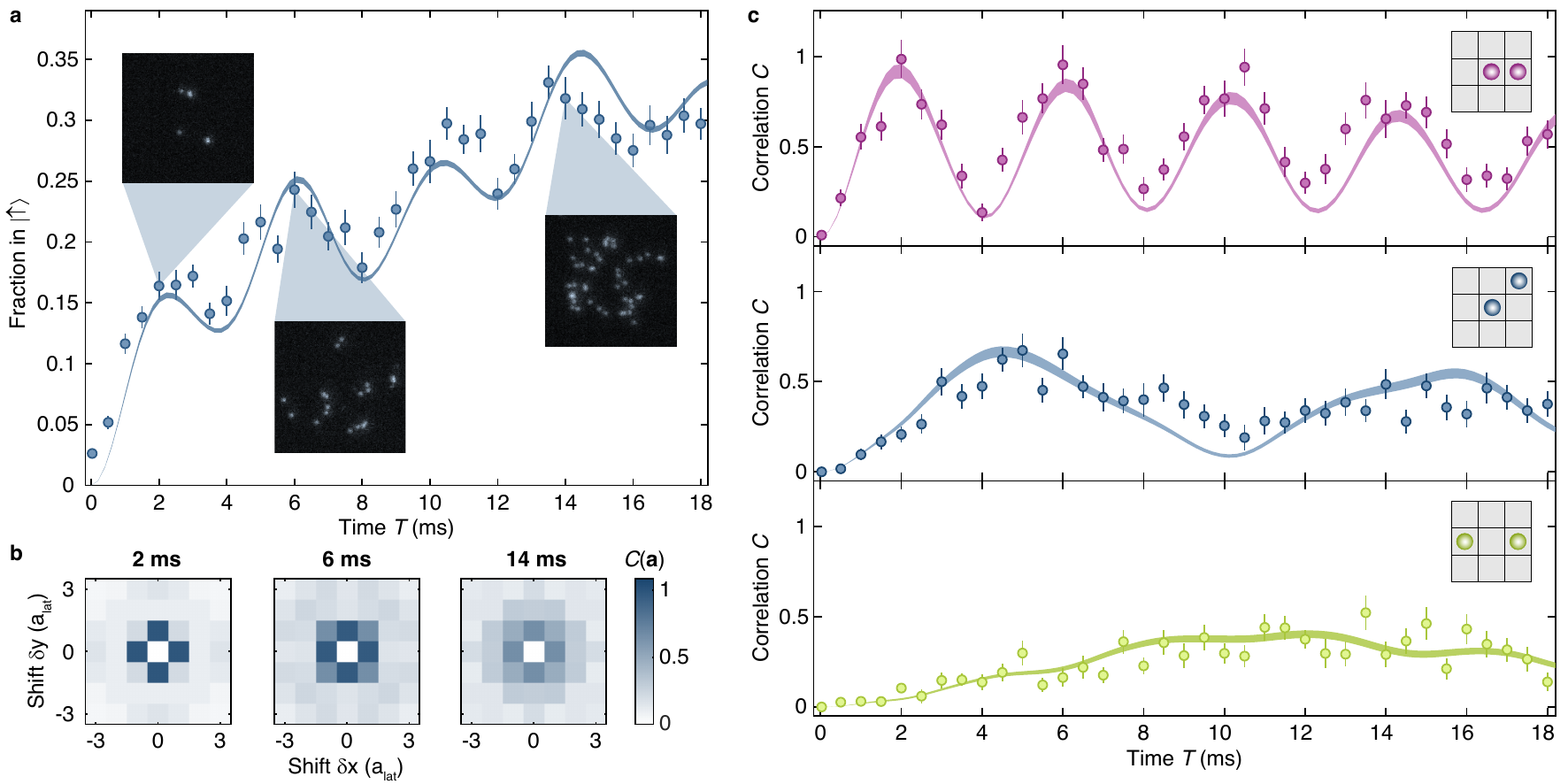}
\caption{\label{fig:quench} \textbf{Correlation dynamics in an XY spin model.} \textbf{a}, Fraction of molecules in $\upstate$ versus time. Insets show representative fluorescence images from \SI{2}{\milli\second} (left), \SI{6}{\milli\second} (middle) and \SI{14}{\milli\second} (right). \textbf{b}, Representative correlation matrices corresponding to the precession times for the inset images in \textbf{a}. All correlations are averaged along the lattice symmetry axes. \textbf{c}, Line plots showing the correlations versus precession time for the specific displacements shown in the inset lattice diagrams. Top: nearest-neighbor correlations. Middle: next-nearest-neighbor correlations. Bottom: next-next-nearest neighbor correlations.  50 images were collected for each precession time. The correlations were averaged along the lattice symmetry axes. All shaded bands are theory predictions of the dynamics from exact diagonalization of the dipolar XY model. The theoretical predictions for the correlations were scaled vertically to best fit the experimental data for the shown displacements simultaneously (see Methods). Error bars are s.e.m.}
\end{figure*}

Having established that the molecules in the lattice can be modeled as a closed quantum spin system out to long times, we probe the growth of their correlations due to dipolar interactions after a quench. For these experiments, we set the peak molecule filling to 5.4(4)\%, which is deliberately chosen to be lower than our maximum filling shown in Fig.~\ref{fig:qgm} in order to slow down the thermalization timescale compared to the  nearest-neighbor interaction timescale. Under these conditions, we expect to observe coherent oscillations in the dynamics of various observables. For simplicity, we start with the magnetic field, which sets the quantization axis for the molecules, pointing perpendicular to the 2D lattice plane so that the dipolar interactions between the molecules are isotropic. The quench is initiated with a global $\pi/2$ rotation to transfer the molecules from $\upstate$ to $\ket{+X}=(\downstate + \upstate)/\sqrt{2}$. The interacting molecules then evolve for a time $T$, with an odd number of spin echo pulses introduced to mitigate residual effective fields (see Methods). Finally, we rotate the spins into the measurement basis with a final $\pi/2$-pulse, which is $\SI{180}{\degree}$ out of phase with the initial pulse so that in the absence of any many-body interactions all of the molecules are transferred to $\downstate$ and therefore appear dark to our detection scheme. Any molecules in $\upstate$ are then detected and their positions in the lattice are recorded.  

The results of this experiment, along with comparison to numerical simulations, can be seen in Fig.~\ref{fig:quench}. Initially there are a negligible number of molecules in $\upstate$, but by \SI{18}{\milli\second} approximately $30\%$ of the molecules are in $\upstate$. However, this overall loss of magnetization during the evolution of the system is not monotonic, as clear oscillations in the molecule number can be seen in Fig.~\hyperref[fig:quench]{3a} in good agreement with the theoretically expected frequency of $|V(\mathbf{e})|/2h = \SI{241}{Hz}$, where $\mathbf{e}=(1,0)$, corresponding to the interaction energy between spins on neighboring sites. While such oscillations have been previously observed~\cite{yan2013observation}, our site-resolved measurements additionally allow us to extract the lattice-averaged correlation function~$C(\mathbf{a},T) = \mathcal{N}\sum_\mathbf{r}{(\langle n^\uparrow_\mathbf{r} n^\uparrow_{\mathbf{r}+\mathbf{a}} \rangle - \langle  n^\uparrow_\mathbf{r} \rangle \langle n^\uparrow_{\mathbf{r}+\mathbf{a}} \rangle)}$, yielding a more detailed characterization of the system's dynamics. Here, $\langle .\rangle$ denotes averaging over the quantum state of the many-body spin system as well as classical realizations of the dilute lattice filling (a disorder average). The normalization factor is $\mathcal{N} = (N_s\expval{\rho^2}_\text{lat})^{-1}$, where $\expval{\rho^2}_\text{lat}$ is the lattice average of the square of the filling fraction and $N_s$ is the number of sites in the region used to evaluate the correlator.

At short evolution times the molecules detected in $\upstate$ are predominantly nearby pairs or few-spin clusters, as can be seen in a sample fluorescence image (Fig.~\hyperref[fig:quench]{3a} inset) and the corresponding correlation matrices in Fig.~\hyperref[fig:quench]{3b}. This is because the dynamics of spins that do not have another spin nearby are only governed by the odd number of spin echo pulses on these time scales, so they end up in the undetected state $\downstate$, whereas interactions play an important role in the dynamics of pairs or clusters of nearby spins, so they can evolve into the detected state in a correlated way. For example, the quantum state of an isolated pair of molecules with displacement $\mathbf{a}$ at the end of a Ramsey sequence with evolution time $T$ is $\textrm{sin}(V(\mathbf{a})T/4)\ket{\uparrow\uparrow} -i\textrm{cos}(V(\mathbf{a})T/4) \ket{\downarrow\downarrow}$ up to a global phase factor. This two-molecule entangling-disentangling dynamics is responsible for an oscillatory behavior of the correlations given by $C(\mathbf{a},T)\sim \textrm{sin}^2(V(\mathbf{a})T/4)$ in the limit of vanishing lattice filling and for short evolution times. This limiting case underlies the correlation oscillations in the data in Fig.~\hyperref[fig:quench]{3c}. These oscillations are observed to damp slowly both in the data and in exact diagonalization calculations using the dipolar XY model. Since the numerics are performed for a closed quantum system, we understand the damping to be due to corrections to the simple two-spin picture above: pairs of nearby spins contribute dominantly to the dynamics and couplings from those pairs to other spins that are farther away, due to the long-range interactions, play an increasingly important role as time goes on.

\section{Tunable Spatial Anisotropy}
\begin{figure*}[th]
\includegraphics[width=\textwidth]{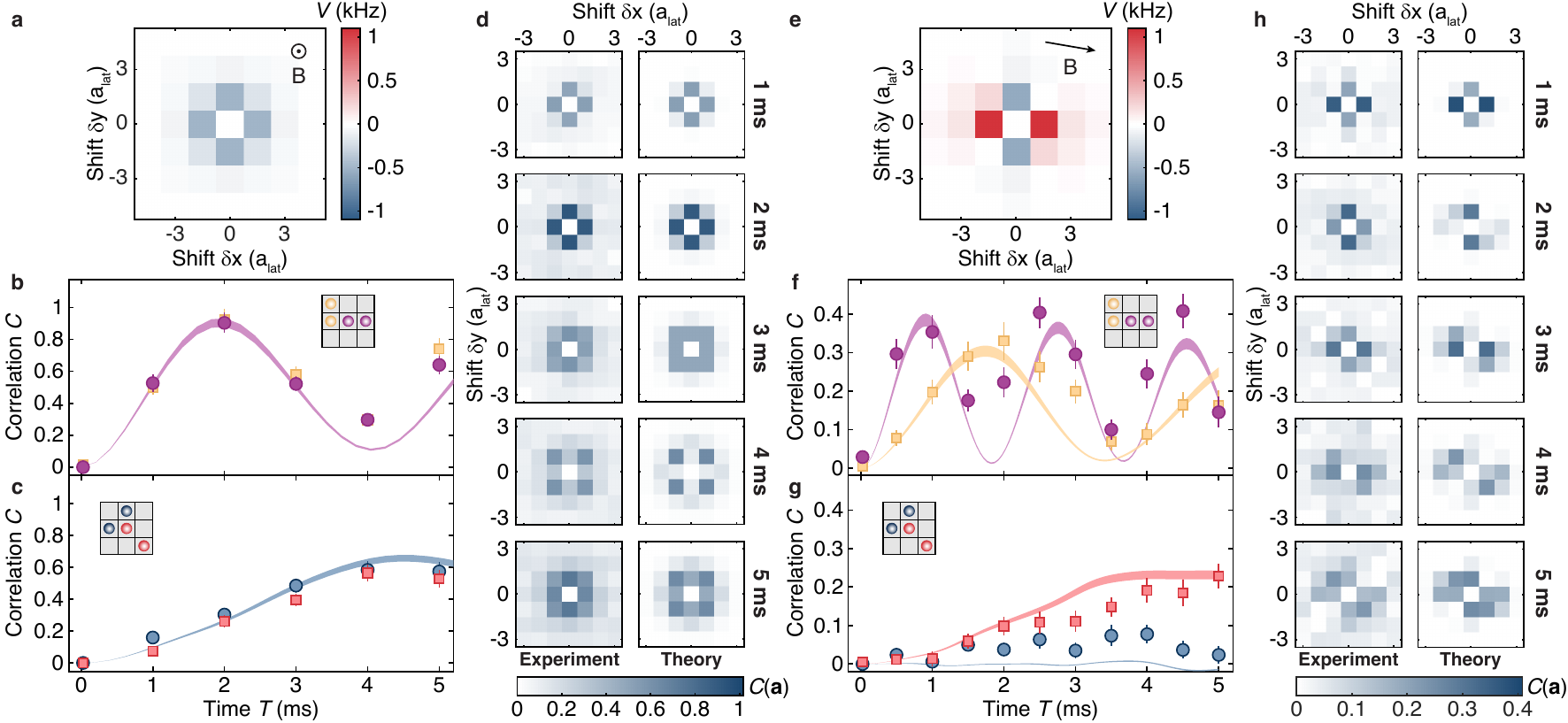}
\caption{\label{fig:isoaniso} \textbf{Tunable spatial anisotropy of the dipolar interactions.} \textbf{a}, Interaction potential between the molecules when the magnetic field is perpendicular to the 2D lattice plane (see Methods). \textbf{b}, Nearest-neighbor correlations along the vertical (yellow) and horizontal (purple) lattice axes for different evolution times for the isotropic configuration. \textbf{c}, Next-nearest-neighbor correlations along $y=x$ (blue) and $y=-x$ (red) for different evolution times for the isotropic configuration. \textbf{d}, Full correlation matrices obtained from 200 experimental iterations for each evolution time (left) compared to theory predictions from exact diagonalization (right). \textbf{e}, Interaction potential between the molecules when the magnetic field is in the 2D lattice plane and oriented 9$^\circ$ from the lattice axes (see Methods). \textbf{f}, Nearest-neighbor correlations along the vertical (yellow) and horizontal (purple) lattice axes for different evolution times for the anisotropic configuration. \textbf{g}, Next-nearest-neighbor correlations along $y=x$ (blue) and $y=-x$ (red) for different evolution times for the spatially anisotropic configuration. \textbf{h}, Full correlation matrices obtained from 50 experimental iterations for each evolution time (left) compared to theory predictions from exact diagonalization (right). All shaded bands are theory predictions from exact diagonalization of the dipolar XY model. Inset lattice diagrams show the specific site displacements used to calculate the correlations. The theoretical predictions for the correlations were scaled vertically to best fit the experimental data for the shown displacements simultaneously (see Methods). Error bars are s.e.m.}
\end{figure*}

Another defining characteristic of the dipole-dipole interaction is its spatial anisotropy.  By performing the same quench dynamics experiment but with the magnetic field in the plane and tilted \SI{9}{\degree} with respect to the lattice axes, we can clearly see the effect of the anisotropic interaction potential in the dynamics. The resulting correlations are shown in  Fig.~\ref{fig:isoaniso}, alongside the full correlation matrices for the isotropic configuration (discussed in the last section), to highlight the change from isotropic to anisotropic correlations. For example, in contrast to the correlations shown for the isotropic case, the correlations along the $x$-axis grow faster than the correlations along the $y$-axis, as expected since the interaction energy is a factor of ${\sim}\,$2 stronger along this axis. In addition, the interaction potential is zero for $\theta=\SI{54.7}{\degree}$, which falls near the $y\,{=}\,x$ diagonal, where $x$ and $y$ are the lattice axes. 
Therefore we expect the correlations to be highly suppressed along that direction. We observe this in the experimental correlation data in Fig.~\hyperref[fig:isoaniso]{4h}, where it can be seen at 4 and \SI{5}{\milli\second} that the $y \,{=} -x$ diagonal has strong correlations but there are negligible correlations for the $y\,{=}\,x$ diagonal. 

We note that similar techniques have been applied to Rydberg-dressed atoms to map out the shape of the long-range interaction potential using spin correlations~\cite{zeiher2016many}. Our correlation data demonstrate the ability to easily tune the molecular interaction anisotropy with the direction of the magnetic field, which will be useful in future studies of frustrated spin systems. 

 \section{Floquet engineering of an XXZ model}
 \begin{figure*}[th]
\includegraphics[width=\textwidth]{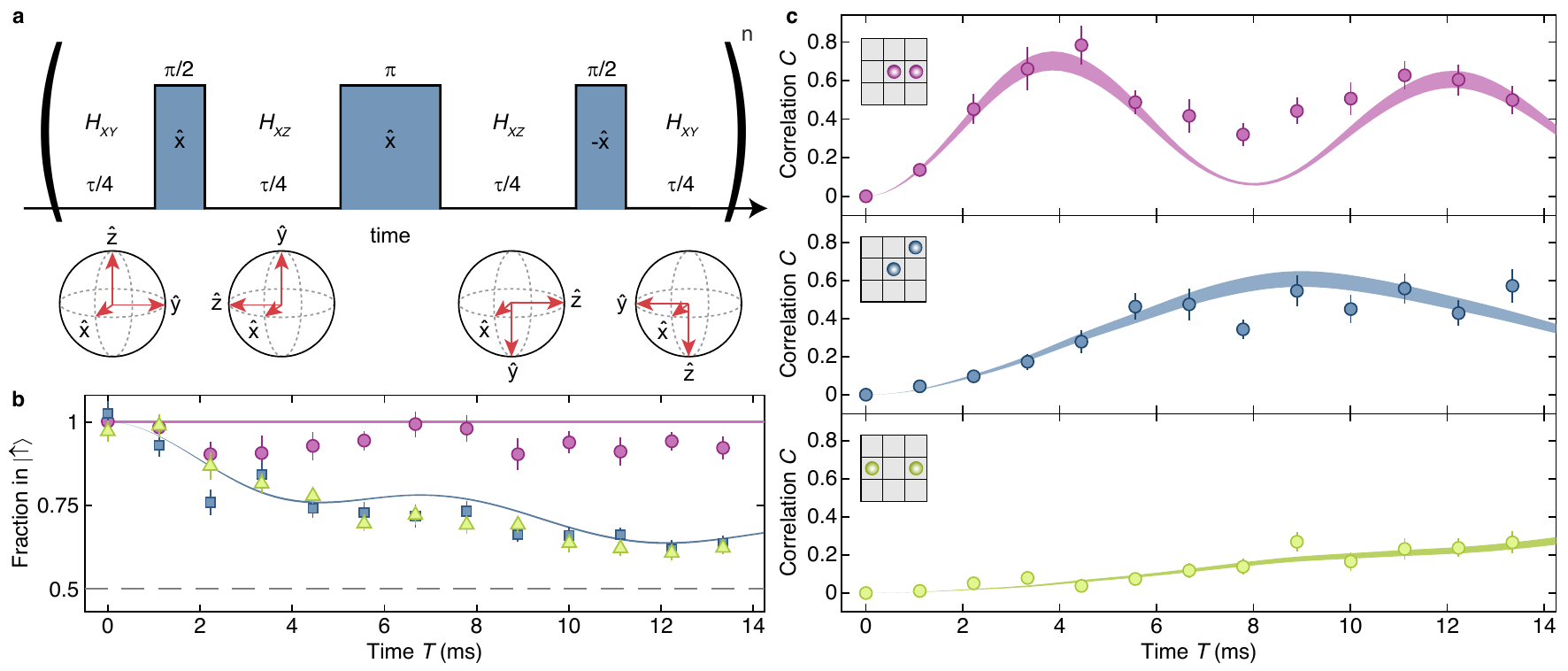}
\caption{\label{fig:Floquet} \textbf{Floquet engineering of an anisotropic Heisenberg model.} \textbf{a}, Pulse sequence used for an individual Floquet cycle, which is repeated $n$ times. Each microwave pulse separates an equal time segment $\tau/4$. \textbf{b}, Magnetization dynamics for three different initial states. Purple: $\ket{+X}$ initial state. Blue: $\ket{+Y}$ initial state. Green: $\ket{+Z}$ initial state. \textbf{c}, Correlation dynamics for a $\ket{+Z}$ initial state. Top: nearest-neighbor correlations. Middle: next-nearest-neighbor correlations. Bottom: next-next-nearest neighbor correlations. All shaded bands are theory predictions of the dynamics from exact diagonalization of the XXZ Hamiltonian. Inset lattice diagrams show the specific site displacements used to calculate the correlations. The theoretical predictions for the correlations were scaled vertically to best fit the experimental data for the shown displacements simultaneously (see Methods). Error bars are s.e.m.}
\end{figure*}

Although the XY model is the natural spin Hamiltonian implemented in our polar molecule system, one can go beyond the resonant spin-exchange term to realize more general systems, such as the XXZ or XYZ models, with different symmetries or even the absence of unitary symmetries.
In ultracold atoms, transport properties were studied in the context of a Heisenberg model with fully tunable, short-ranged spin-spin couplings~\cite{jepsen2020}.
For polar molecules, an Ising term of arbitrary strength can be introduced by either the application of a dc electric field~\cite{hazzard2013far,bohn2017cold} or microwave Floquet engineering.
The essence of the latter technique is to modulate the Hamiltonian in time at a rate much faster than the interaction timescale, so that the time-averaged Hamiltonian is different from the original Hamiltonian. This technique was recently used in the context of Rydberg atom quantum simulators~\cite{geier2021floquet,scholl2022}, and can be used in a variety of many-body spin systems~\cite{choi2020robust}.  Here we use rapid periodic $\pi/2$ rotations to realize an effective XXZ Hamiltonian. An additional $\pi$-pulse in the center of the cycle serves as a spin echo to cancel the dephasing due to any effective fields without affecting the many-body interactions. The total Floquet cycle can be seen in Fig.~\hyperref[fig:Floquet]{5a}. The effect of each Floquet cycle is to evolve the molecules for an equal amount of time under $H_{XY}$ and $H_{XZ}$, time averaging to 
 \begin{equation}
    H_{\rm XYY} =  \sum_{i > j} V(\mathbf{r}_i-\mathbf{r}_j) \left(S^x_iS^x_j + \frac{1}{2}(S^y_iS^y_j + S^z_iS^z_j)\right)
    \label{eq:Floquet}
\end{equation} which is equivalent to the XXZ  model with a permutation of the axis labels.

One important feature of the Hamiltonian in equation~(\ref{eq:Floquet}) is that the magnetization along $x$ is conserved, whereas that along $y$ and $z$ are not, due to the $U$(1) symmetry corresponding to rotations about the $x$-axis. Therefore, a key test of engineering the correct effective Hamiltonian is to verify this spin conservation. We prepare three distinct initial states, corresponding to all of the spins pointing along $x$, $y$ or $z$, and then subject the spins to the same Floquet Hamiltonian using the pulse sequence shown in Fig.~\hyperref[fig:Floquet]{5a}.  We set the Floquet evolution time to $\tau = \SI{1}{\milli\second}$ and use a $\pi$-pulse duration of \SI{56.2}{\micro\second} so that each Floquet cycle is short compared to the interaction period of the original XY Hamiltonian seen in Fig.~\ref{fig:quench}. The density of the molecules is 8.1(8)\%, and the quantization axis is perpendicular to the 2D lattice so that the interactions are isotropic in the plane. Finally, we measure the total number of spins that remain in the initial state for different evolution times. For the initial states $\ket{+Y}$ and $\ket{+Z}$, an additional $\pi$-pulse is added at the end of the evolution time to return the coordinate system of the Bloch sphere to its initial orientation before performing measurements. 

The results of this experiment are shown in Fig.~\hyperref[fig:Floquet]{5b}. The magnetization of molecules initially prepared in the $\ket{+X}$ state remains approximately unchanged throughout the dynamics, whereas that of molecules prepared in the $\ket{+Y}$ and $\ket{+Z}$ states decays towards the unpolarized value of $1/2$. In addition, the data is $Y$-$Z$ symmetric, as expected from equation~(\ref{eq:Floquet}). Comparing the experimental data to exact diagonalization calculations using an XXZ  model shows reasonable agreement, indicating that the time-averaged Hamiltonian is an appropriate descriptor of the observed dynamics.

Finally, we can measure the molecule correlations as a function of time to test whether the oscillation frequency changes for the Floquet dynamics compared to the original XY Hamiltonian. Starting with a peak molecule filling of 3.1(4)\%, we prepare all of the spins in $\upstate$, quench the system to evolve under the Floquet Hamiltonian and then measure the correlations between the molecules after each Floquet cycle. By comparing these data in Fig.~\hyperref[fig:Floquet]{5c} to the XY experiment in Fig.~\hyperref[fig:quench]{3c}, we can see that the oscillation frequency for the nearest-neighbor correlations has been halved, as expected for the desired Floquet Hamiltonian. 
To conclude, we demonstrate that microwave Floquet engineering of polar molecule systems enables the study of spin-anisotropic Heisenberg models, further enriching the quantum simulation toolbox in this system.

\section{Outlook}

The experiments described above demonstrate the powerful capabilities of ultracold molecules to study dynamics in the context of quantum magnetism, and open up new avenues for studying quantum physics with ultracold molecules more broadly. The ability to detect correlations with single-site resolution could be used in other experiments to distinguish different equilibrium phases of matter, such as a rotational superfluid that should emerge by adiabatic preparation of the ground state of the XY Hamiltonian even at current demonstrated filling fractions~\cite{kwasigroch2014bose}. 
The single-site resolution also enables optical manipulation of the underlying potential at the smallest relevant length scale, allowing the initialization of arbitrary configurations to study impurity physics or transport dynamics~\cite{gross2021quantum}. In addition, while so far we have demonstrated interactions between molecules that are frozen in the lattice, complex phases of matter are also expected to emerge upon allowing the molecules to tunnel throughout the lattice. The interplay between the dipolar interactions and the kinetic energy of the molecules is predicted to yield a rich phase diagram consisting of superfluid and supersolid phases, as well as fractional Mott insulators~\cite{barnett2006quantum,buchler2007strongly,capogrosso2010quantum}. Finally, the capabilities demonstrated here to measure correlations between interacting polar molecules may prove useful in the context of quantum computing~\cite{DeMille2002,ni2018dipolar}, where they could be used to verify the creation of Bell and GHZ states as well as for quantum error correction.

\textbf{Acknowledgements:} We would like to thank Elmer Guardado-Sanchez and Geoffrey Zheng for experimental assistance. This work was supported by the NSF (grant no. 1912154) and the David and Lucile Packard Foundation (grant no. 2016-65128). L.C. was supported by the NSF Graduate Research Fellowship Program.  D.A.H. and A.M. were supported in part by NSF QLCI grant OMA-2120757.

\textbf{Author contributions:} W.S.B. and D.A.H. conceived the study and supervised the experiment. L.C., J.S.R., R.R. and Z.Z.Y. performed the experiments and the data analysis. A.M. and S.C. performed the numerical calculations. All authors contributed to the manuscript.

\textbf{Competing interests:} The authors declare no competing interests.

\section{Methods}
\setcounter{figure}{0}
\setcounter{equation}{0}
\setcounter{section}{0}
\renewcommand{\thefigure}{S\arabic{figure}}
\renewcommand{\thetable}{S\arabic{table}}
\renewcommand{\theequation}{S\arabic{equation}}
\renewcommand{\theHfigure}{S\arabic{figure}}
\renewcommand{\theHtable}{S\arabic{table}}
\renewcommand{\theHequation}{S\arabic{equation}}

\subsection{Molecule preparation}
 The molecules are created by first loading a doubly-degenerate mixture of Na and Rb atoms into a single plane of a 3D optical lattice. The 2D lattice in the $x$-$y$ plane has a \SI{752}{\nano\metre} spacing and the vertical lattice has a \SI{3.8}{\micro\metre} spacing. We then sweep the magnetic field across an interspecies Feshbach resonance at \SI{347.6}{G} to form weakly-bound molecules. Each lattice site that contains one Na and one Rb atom forms a weakly-bound molecule, whereas sites with more than one atom of each species are emptied by three-body loss processes. Remaining Na and Rb atoms that do not form molecules are removed with resonant light pulses. Details of the optical potentials and Feshbach molecule creation process can be found in Ref.~\cite{rosenberg2021observation}.

After forming the Feshbach molecules, we ramp the magnetic field to \SI{335}{G} and transfer the molecules to the electronic and rovibrational ground state via stimulated Raman adiabatic passage (STIRAP). Following the procedure described in~\cite{guo2016creation}, we use two external-cavity diode lasers (Toptica DL Pro) with wavelengths of \SI{770}{\nano\metre} and \SI{1248}{\nano\metre}, locked to a common high-finesse ultra-low expansion cavity  (Stable Laser Systems) via the Pound-Drever-Hall (PDH) technique. The cavity free-spectral range is \SI{1.5}{\giga\hertz} and the finesse is \num[group-separator = {,}]{34000} at \SI{770}{\nano\metre} and \num[group-separator = {,}]{43000} at \SI{1248}{\nano\metre}. We use a fiber electro-optic modulator (EOM) for each laser (EOSpace) to tune the frequency of the laser relative to that of a fixed cavity mode, as well as to generate sidebands for the PDH lock. The laser linewidth is narrowed by active feedback of the diode current, whereas slow drifts in the frequency are compensated with a separate feedback loop stabilizing the laser piezo. Each laser uses a FALC 110 (Toptica) laser locking module for both feedback loops. The Rabi frequencies are $2\pi\times\SI{0.70(4)}{\mega\hertz}$ and $2\pi\times\SI{0.9(1)}{\mega\hertz}$ for the \SI{1248}{\nano\metre} and \SI{770}{\nano\metre} transitions respectively, and we typically achieve 93.9(3)\% STIRAP efficiency. The molecules are prepared in the stretched hyperfine state $\ket{m_{I,\mathrm{Na}},m_{I,\mathrm{Rb}}} = \ket{3/2,3/2}$. To deliberately dilute the density of the molecules for some of the experiments presented in the paper, we reduce the one-way STIRAP efficiency by decreasing the optical power in the \SI{770}{\nano\metre} beam to reach the desired filling fraction. 

The 2D lattice depth for all of the experiments was 49~$E_R$ with the exception of the coherence data for which the lattice depth was 34~$E_R$. Here $E_R=h^2/8m a_\mathrm{lat}^2$ where $m$ is the molecule mass and $a_\mathrm{lat}=\SI{752}{\nano\metre}$. At both depths, the tunneling of the molecules in the lattice was negligible over the course of the experiment.

\subsection{Rotational states}

\begin{figure}[tb]
\includegraphics[width=\columnwidth]{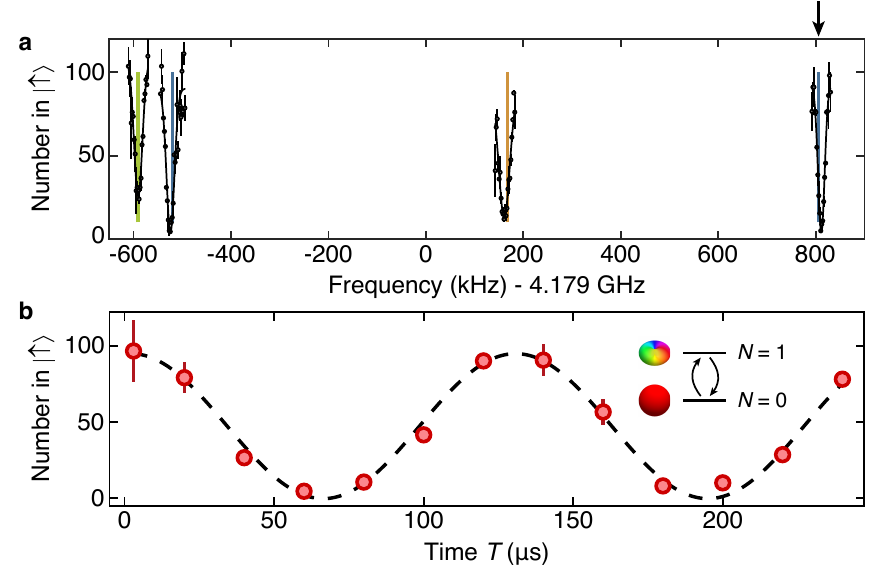}
\caption{\label{fig:spectroscopy} \textbf{Molecular rotational states.} \textbf{a}, Rotational state spectrum measured at \SI{60}{G}. Green, blue and orange lines are the theoretical predictions using molecular parameters in Refs.~\cite{guo2018high,lin2021anisotropic} for the transitions to states using $\pi$, $\sigma^-$, and $\sigma^+$ polarization, respectively. The transition on the far right, indicated by the black arrow, is the $\upstate$ to $\downstate$ transition. \textbf{b}, Sample Rabi oscillation between $\upstate$ and $\downstate$. The extracted Rabi frequency for this data is $2\pi\times\SI{7.8(1)}{\kilo\hertz}$. Top right inset: schematic of the two-level system undergoing Rabi oscillations. Error bars are s.e.m.}
\end{figure}

In order to simulate an effective spin-1/2 system, it is necessary to couple two rotational states that are well-separated in energy from other molecular states and that are as insensitive as possible to noise from variations in magnetic or optical fields. To achieve this condition, we use microwaves resonant with the transition between the ground and first excited rotational state of the molecules. In general the molecule can be in a superposition of states labeled $\ket{N,m_N,m_{I,\mathrm{Na}},m_{I,\mathrm{Rb}}}$, where $N$ is the rotational angular momentum of the molecule, $m_N$ is its projection onto the quantization axis set by the external magnetic field, and $m_{I,\mathrm{Na(Rb)}}$ is the projection of the nuclear spin of the Na (Rb) atom. A representative microwave spectrum with several of the rotational transitions is shown in Fig.~\ref{fig:spectroscopy}, along with a representative Rabi oscillation between the specific pair of states used in this work.

\begin{figure}[htb]
\includegraphics[width=\columnwidth]{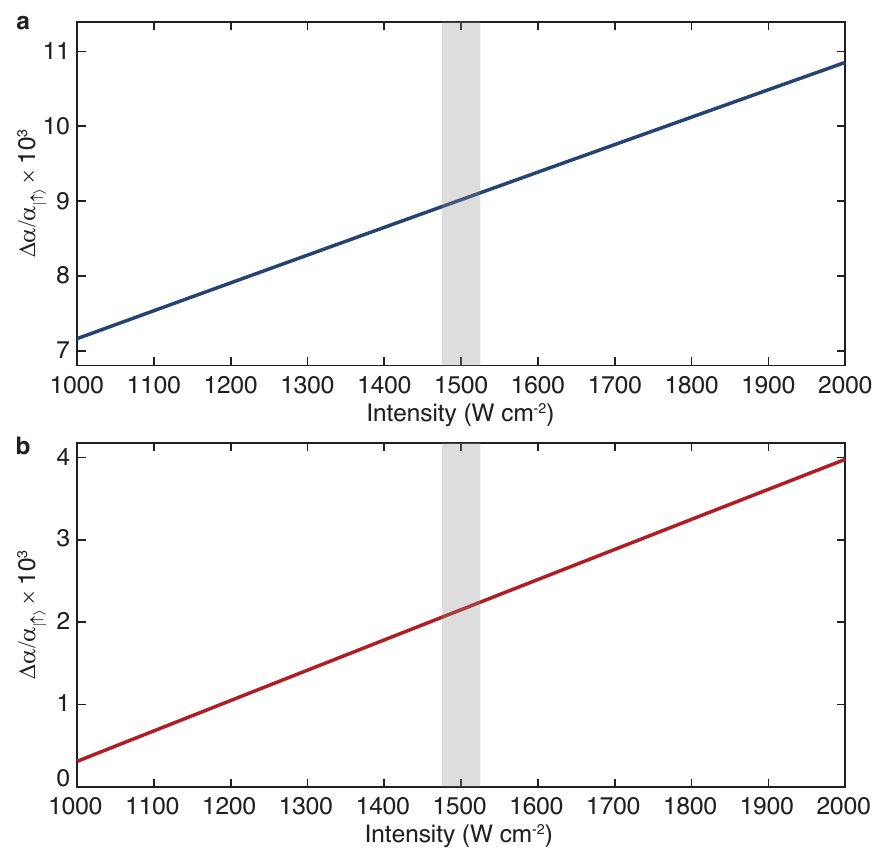}
\caption{\label{fig:alpha} \textbf{Differential polarizabilities between $\upstate$ and $\downstate$ versus trapping light intensity.} \textbf{a}, In the isotropic configuration, $B=\SI{60}{G}$, and the angle between the light's electric field and the quantization axis is \SI{0}{\degree}. The intensity varies by $\sim4$\% over the cloud, denoted by the grey shading. \textbf{b}, In the anisotropic configuration, $B=\SI{4.1}{G}$, and the angle is \SI{90}{\degree}.}
\end{figure}

One important source of single-particle decoherence is the differential ac polarizability $\Delta\alpha=\alpha_{\upstate}-\alpha_{\downstate}$~\cite{Kotochigova2010}. This causes an unwanted shift of microwave transition frequencies across the trap due to the spatially-varying intensity profile of the trapping light.
To mitigate this, we choose two specific magnetic field configurations that produce near-``magic" trapping conditions, where $\Delta\alpha/\alpha_{\upstate}<0.01$. For the experiments with isotropic interactions, the magnetic field is \SI{60}{G}, whereas for anisotropic interactions the magnetic field is \SI{4.1}{G}. While a higher field would have been preferable for the latter case to increase the energy splittings to other states outside of the Hilbert space of the spin system, our coil geometry limits the maximum magnetic field that can be applied in the lattice plane. Our technique makes use of the fact that at zero electric field and weak magnetic fields, $m_N$ is not a good quantum number for the molecular Hamiltonian due to strong hyperfine couplings, especially the nuclear quadrupole moment coupling in $N=1$.  This coupling gives a specific $N=1$ state an admixture of other hyperfine states, which can be leveraged to match the $\downstate$ polarizability to the $\upstate$ polarizability. The three main components of $\downstate$ are  
\begin{align}
     \ket{\downarrow} &\approx 0.688 \ket{1,-1,3/2,3/2} - 0.569\ket{1,0,3/2,1/2} \nonumber \\ &\quad +0.448\ket{1,1,3/2,-1/2}
\end{align}
at \SI{60}{G}, and 
\begin{align}
     \ket{\downarrow} &\approx 0.715 \ket{1,-1,3/2,3/2} - 0.562\ket{1,0,3/2,1/2} \nonumber \\ &\quad +0.413\ket{1,1,3/2,-1/2}
\end{align}
at \SI{4.1}{G}, where the admixtures are calculated using molecular parameters from Refs.~\cite{guo2018high,lin2021anisotropic}. 
Calculated differential polarizabilities are shown in Fig.~\ref{fig:alpha}. Over the spatial extent of the cloud, the microwave transition frequency varies by $<\SI{20}{\Hz}$ due to the inhomogeneous intensity profile. This is consistent with the observed coherence decay time in the absence of a spin echo pulse (Fig.~\ref{fig:coherence}).

\subsection{Microwave control}

\begin{figure*}[tb]
\includegraphics[width=\textwidth]{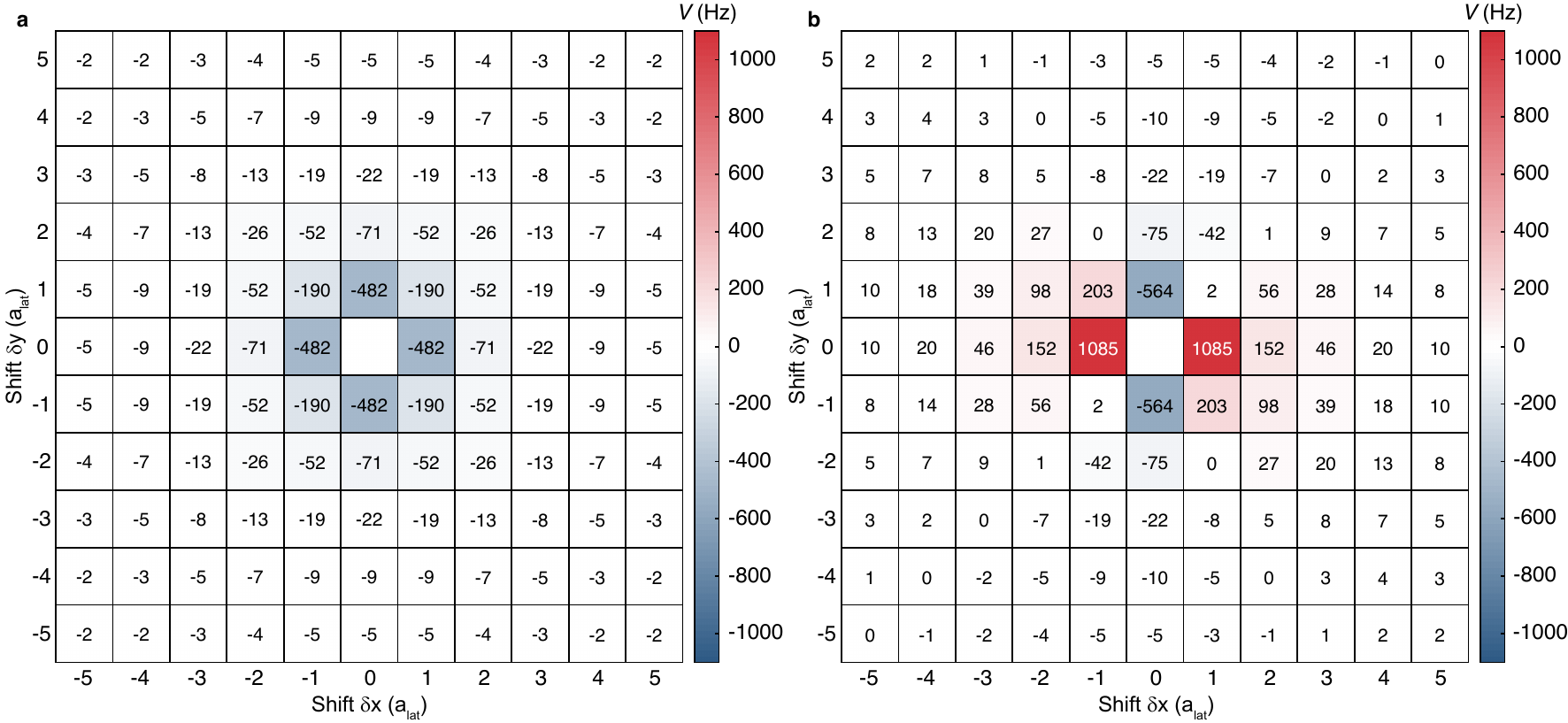}
\caption{\label{fig:Jcalc} \textbf{Spin-exchange coupling.} The values of $V(\mathbf{a})$ calculated for the isotropic (\textbf{a}) and anisotropic (\textbf{b}) cases for different separations in $x$ and $y$.}
\end{figure*}

After transferring the molecules to the rovibrational ground state, transitions to the first excited rotational state are induced using microwaves near \SI{4.18}{\giga\hertz}. The microwaves are generated by mixing a local oscillator at \SI{4.13}{\giga\hertz} provided by an analog signal generator (Agilent E8257C) with an intermediate frequency provided by an arbitrary waveform generator (Keysight 33600A). The intermediate frequency is \SI{50.13747}{\mega\hertz} when the magnetic field is \SI{60}{G}, and \SI{50.08300}{\mega\hertz} when the magnetic field is \SI{4.1}{G}. The mixed signal is then amplified (Mini-Circuits ZHL-5W-63-S+) before being sent to a homebuilt helical antenna mounted underneath the vacuum chamber. We program the Keysight arbitrary waveform generator to set the duration, amplitude, and phase necessary for each pulse within the experiments, and keep the Agilent signal generator at fixed frequency, amplitude and phase. The oven-controlled crystal oscillator inside the Agilent signal generator is used as a common \SI{10}{\mega\hertz} clock for both instruments.

\subsection{Pulse sequences}
We use a microwave Rabi frequency of \SI{9}{\kilo\hertz} for all of the experiments with spatially isotropic interactions. This frequency is chosen to be large compared to the intermolecular interactions ($<\SI{1}{\kilo\hertz}$) but small compared to the splitting between the different hyperfine states to avoid driving transitions outside of the effective two-level system. However for the spatially anisotropic correlation measurement we decrease the Rabi frequency to \SI{4}{\kilo\hertz} to minimize Fourier broadening from the microwave pulses, since the hyperfine states are more closely spaced at the lower magnetic field that we use for these experiments. 

In the Ramsey spectroscopy experiments shown in Fig.~\ref{fig:coherence}, we observe a slow drift in the phase of the fringe in addition to a decay in the fringe amplitude. This phase drift is not expected from the desired Hamiltonian in equation~(\ref{eq1}), and can be caused by a combination of the interactions and the inhomogeneous light shifts, or an unknown time-varying field present in the lab.  We observe that adding multiple $\pi$-pulses does not affect the decay rate of the fringe amplitude, but it does remove the phase drift. Therefore, for all of the dynamics experiments in Figs.~\ref{fig:quench} and \ref{fig:isoaniso}, we use one $\pi$-pulse for any data collected with less than \SI{10}{\milli\second} of evolution time, where the phase shift is negligible. For data collected after \SI{10}{\milli\second} of evolution time, more than one $\pi$-pulse is required to mitigate the phase shift, so we use three $\pi$-pulses during the evolution time, deliberately keeping the number of $\pi$-pulses odd. In addition, for each experiment we synchronize the start of the first microwave pulse with the ac line to ensure that every experiment is performed with the same background magnetic field conditions.

\subsection{Calculation of spin-exchange coupling}\label{sec:Jcalc}
Each term $V(\mathbf{r}_i - \mathbf{r}_j)$ in equation~(\ref{eq1}) can be calculated numerically.
With $\upstate$ and $\downstate$ as the pseudospin-1/2 system, the spin-exchange interaction experienced by two molecules with wavefunctions $\psi_1(\mathbf{r}_i)$ and $\psi_2(\mathbf{r}_j)$ on lattice sites $\mathbf{r}_i,\mathbf{r}_j$ is given by
\begin{align}
    &V(\mathbf{r}_i - \mathbf{r}_j) =\\ &J 
    \int d\mathbf{r}_i d\mathbf{r}_j |\psi_1(\mathbf{r}_i)|^2 |\psi_2(\mathbf{r}_j)|^2 \left(\frac{1-3\cos^2\theta(\mathbf{r}_i - \mathbf{r}_j)}{|\mathbf{r}_i-\mathbf{r}_j|^3}\right) \nonumber
\end{align}

where \begin{equation}
    J = |\braket{1,-1,3/2,3/2}{\downarrow}|^2 \left(- \frac{1}{4\pi\epsilon_0 a_\mathrm{lat}^3} \frac{d^2}{3} \right). 
\end{equation}
Here, $d=\SI{3.3}{D}$ is the permanent body-frame dipole moment~\cite{vexiau2017dynamic}, $\epsilon_0$ is the vacuum permittivity, and $\braket{1,-1,3/2,3/2}{\downarrow}$ is the wavefunction overlap between the target $\downstate$ state in the $N\,{=}\,1$ manifold and the $\ket{1,-1,3/2,3/2}$ state, leading to $J/h = \SI{610}{\hertz}$ at \SI{60}{G} and $J/h = \SI{659}{\hertz}$ at \SI{4.1}{G}. The dominant source of uncertainty in $J/h$ is a difference of $\sim\SI{0.1}{D}$ between theoretical~\cite{vexiau2017dynamic} and experimental~\cite{guo2016creation} values of $d$, leading to corrections of $\sim6$\% in $J/h$ for the isotropic and anisotropic cases. Uncertainties in $a_\mathrm{lat}$ lead to corrections of $<3\%$. The molecules occupy the ground center-of-mass state of their respective lattice sites, so their wavefunctions are approximated by the 3D harmonic oscillator ground state wavefunctions given by $\psi(\mathbf{r}_i) = \psi(\mathbf{x}_i)\psi(\mathbf{y}_i)\psi(\mathbf{z}_i)$.
The axial ($z$) and radial ($x,y$) trap frequencies are \SI{2}{\kilo\hertz} and \SI{9}{\kilo\hertz}, respectively.

The resulting calculations are shown in Fig.~\ref{fig:Jcalc} and in Fig.~\ref{fig:isoaniso} of the main text.

\subsection{Numerical simulations}

Theoretical data in the figures were generated using exact numerical diagonalization of the XY and XXZ Hamiltonians for samples of $N=12$ spins randomly placed on a square lattice of size $L\times L$ with periodic boundaries. The spin-spin interactions for a given displacement are set as described in the previous section, such that the large-$r$ form of the Hamiltonians match onto equations~(\ref{eq1}) and (\ref{eq:Floquet}) with $|J|/h = \SI{600}{\hertz}$. To accommodate long-range interactions and periodic boundaries, the interaction between two spins is set according to the shortest displacement between the spins. Each data point shown for comparison to theory is generated from 2500 samples. These samples are meant to model local patches of the much larger experimental lattice. Due to the experimental run times being moderate in units of the interaction time for typical spins, simulating a larger number of spins is not necessary. Integer $L$ is chosen for a sample such that $N/L^2$ is closest to a target density. 

The density profile of molecules in the experimental lattice is nonuniform. To model this in the simulations we let the density of our samples be a random variable. Lower (higher) density samples represent patches of the experimental lattice that are farther from (closer to) the center. We assume a radial density profile for the experiment of the form
\begin{align}
    \rho(r) = \rho_\mathrm{max} \left[ 1-\left(\frac{r}{r_\mathrm{max}}\right)^2 \right],
\end{align}
where $\rho_\mathrm{max}$ is the peak filling at the center of the lattice and $r_\mathrm{max}$ is the distance from the center at which the density becomes negligible. None of our theoretical results actually depend on $r_\mathrm{max}$, and the peak densities for each experiment are given in the main text. Sampling patches of $N$ spins with density $\rho$ from such a density profile corresponds to sampling a radial location according to the probability density $p(r)\propto r \rho(r)$. The factor of $r$ is because the amount of lattice at radius $r$ is $\propto r$, and the factor of $\rho(r)$ is because a patch with a predetermined number of spins $N$ covers area $\propto 1/\rho(r)$. Now instead of sampling the position of the patch in the lattice, we change variables and sample the corresponding density of the patch. This yields a probability density $p(\rho) \propto \rho$ for $\rho\le \rho_\mathrm{max}$.

Correlation functions are computed in a way that is similar to the analysis of experimental data: For the experimental data we first compute the quantum-and-disorder-averaged values of $n^\uparrow_\mathbf{r} n^\uparrow_{\mathbf{r}+\mathbf{a}}$ and $n^\uparrow_\mathbf{r}$, then form the position-dependent correlation function $\langle n^\uparrow_\mathbf{r} n^\uparrow_{\mathbf{r}+\mathbf{a}} \rangle - \langle  n^\uparrow_\mathbf{r} \rangle \langle n^\uparrow_{\mathbf{r}+\mathbf{a}} \rangle$ and average that over the lattice. The lattice average is performed last so as to try to delay mixing data from different densities until after the connected correlation function is formed. Similarly, for the simulation data we first compute the connected correlation function independently for groups of samples with the same density, and then we average that density-dependent correlation function, with the weight in the average given by the total lattice area of the samples at that density. 

Both experimental and simulated correlation functions are scaled in the same way, by the lattice-averaged value of the squared density, which is $\rho_\mathrm{max}^2 / 3$ for the theoretical model. In comparing the theoretical correlations to the experimental data, a single fit parameter is introduced to scale the amplitude of the correlations. The amplitude scale factor is obtained by fitting the experimental data for the displacements in the line plots simultaneously and is always found to be order unity (1.20(4) for Fig.~\ref{fig:quench}, 1.27(7) for Fig.~\hyperref[fig:isoaniso]{4b-d}, 0.48(3) for Fig.~\hyperref[fig:isoaniso]{4f-h}, and 1.12(7) for Fig.~\ref{fig:Floquet}). The deviation from unity may be due to inaccuracies in the modeling of the density profile and slow drifts of the atom number over the course of the data collection.

\bibliography{molecule_spin_correlations}

\end{document}